\documentclass[reprint, amsmath, amssymb, pre]{revtex4-2}

\usepackage[utf8]{inputenc}
\usepackage[colorlinks=true, citecolor=blue, linkcolor=blue, breaklinks=true]{hyperref}
\usepackage{lineno}

\usepackage[capitalise, nameinlink]{cleveref}

\usepackage{graphicx}
\graphicspath{{./main_figs/}{./}}
\DeclareGraphicsExtensions{.pdf,.png}

\begin{document}
\title{Geometric frustration of hard-disk packings on cones}
\author{Jessica H. Sun}
\affiliation{Harvard John A. Paulson School of Engineering and Applied Sciences, Harvard University, Cambridge, MA 02138, USA}
\author{Abigail Plummer}
\affiliation{Princeton Center for Complex Materials,
Princeton University, Princeton, NJ 08540, USA}
\author{Grace H. Zhang}
\affiliation{Department of Physics, Harvard University, Cambridge, MA 02138, USA}
\author{David R. Nelson}
\affiliation{Department of Physics, Harvard University, Cambridge, MA 02138, USA}%
\author{Vinothan N. Manoharan}
\affiliation{Harvard John A. Paulson School of Engineering and Applied Sciences, Harvard University, Cambridge, MA 02138, USA}
\affiliation{Department of Physics, Harvard University, Cambridge, MA 02138, USA}

\date{\today}
\pacs{Valid PACS appear here}


\begin{abstract}
  Conical surfaces pose an interesting challenge to crystal growth: a
  crystal growing on a cone can wrap around and meet itself at different
  radii. We use a disk-packing algorithm to investigate how this closure
  constraint can geometrically frustrate the growth of single crystals
  on cones with small opening angles. By varying the crystal seed
  orientation and cone angle, we find that---except at special
  commensurate cone angles---crystals typically form a seam that runs
  along the axial direction of the cone, while near the tip, a
  disordered particle packing forms. We show that the onset of disorder
  results from a finite-size effect that depends strongly on the
  circumference and not on the seed orientation or cone angle. This
  finite-size effect occurs also on cylinders, and we present evidence
  that on both cylinders and cones, the defect density increases
  exponentially as circumference decreases. We introduce a simple model
  for particle attachment at the seam that explains the dependence on
  the circumference. Our findings suggest that the growth of single
  crystals can become frustrated even very far from the tip when the
  cone has a small opening angle. These results may provide insights
  into the observed geometry of conical crystals in biological and
  materials applications.
\end{abstract}
\maketitle
\section{Introduction}

The growth of a crystal can be frustrated by interactions with a curved
surface such as a spherical or hyperbolic
substrate~\citep{nelson_order_1983, grason_perspective_2016,
  hagan_equilibrium_2021, meldrum_crystallization_2020}. When the
surface has nonzero Gaussian curvature, the frustration stems from
variations in the surface metric, which lead to stretching of the
crystal lattice. This type of geometrical frustration has been well
studied, particularly in colloidal systems~\citep{bausch_grain_2003,
  lipowsky_direct_2005, vitelli_crystallography_2006,
  bowick_dynamics_2007, irvine_pleats_2010,
  irvine_fractionalization_2012, meng_elastic_2014,guerra_freezing_2018,
  singh_observation_2022}.

Less well studied is frustration arising on surfaces with no Gaussian
curvature but on which crystals can form loops, such as cylinders.
Although such surfaces do not stretch the lattice, they can nonetheless
frustrate a crystal by imposing a closure constraint. As observed in
experiments and simulations on colloidal crystals on cylindrical
fibers~\citep{tanjeem_geometrical_2021}, crystals with orientations that
are incommensurate with the closure constraint form seams. These seams,
which are stable on cylinders but not on flat surfaces unconstrained by
periodic boundary conditions, break the translational symmetry of the
crystal.

Here we examine how the closure constraint affects crystallization on a
cone, which, unlike a cylinder, has a spatially varying circumference.
As a consequence, a seam must form with a width that varies in the axial
direction whenever the cone angle does not permit the crystal to wrap
perfectly around the cone (for a triangular lattice, such commensurate
wrappings can be achieved by placing, for example, a 60\textdegree{}
disclination at the cone apex). The seam is similar to a tilt grain
boundary between two misoriented crystals on a flat
substrate~\citep{carraro_grain-boundary_1993, coffman_grain_2008},
except that it is a boundary between the misoriented edges of a
\emph{single} crystal that has wrapped around the cone. This seam can
break both the translational and rotational symmetry of the crystal. We
seek to understand how the closure constraint geometrically frustrates
crystal growth on a cone.

There are few previous studies of crystallization on a cone.
Basin-hopping simulations of colloidal crystals showed that interacting
particles on a cone form seams or scar-like
defects~\citep{miller_chapter_2022, gerrand_2021}. The aim of these
simulations was to understand the defect structure and how it changes
with the cone geometry. An experimental study of an atomic system,
WS$_{2}$, showed that crystals on a cone form a distinct
seam~\citep{yu_tilt_2017}. The cone in this work had a large opening
angle, and the size of the WS$_2$ crystal was orders of magnitude larger
than the size of the atoms. The main aim of this study was to
demonstrate the existence of the seam, which the authors refer to as a
tilt grain boundary.

Our aim is to examine the growth process, and in particular how conical
crystals grow after closure. In contrast to the basin-hopping
simulations~\citep{miller_chapter_2022, gerrand_2021}, which reveal
energy-minimizing crystal structures, we aim to determine how an
out-of-equilibrium growth process leads to disorder. Furthermore, we
focus on particles with short-ranged attractions, as seen in previous
experimental studies of crystallization on a flat
surface~\citep{savage_imaging_2006}, sphere~\citep{meng_elastic_2014},
and cylinder~\citep{tanjeem_geometrical_2021}.

To isolate the consequences of geometrical frustration on the crystal
structure---and avoid complications associated with multiple nucleation
sites and kinetics---we use a greedy algorithm to simulate the idealized
growth of a crystal on a conical surface. Our approach, based on an
algorithm developed by Bennett to study metallic
glasses~\citep{bennett_serially_1972}, is inspired by previous work on
understanding the effects of geometrical frustration in metallic
glasses, and on spherical or hyperbolic
surfaces~\citep{rubinstein_order_1982, rubinstein_dense-packed_1983}.
Because we aim to understand effects in the quasi-two-dimensional
colloidal systems realized experimentally, we simulate disk packings
rather than sphere packings to simplify the computation.

Our simulation is designed to model the slow, reaction-limited growth of
a single crystal from a fixed nucleus. Briefly, we initialize the
simulation with three disks placed in a triangular configuration with a
defined orientation and position on the surface (\cref{fig:geometry}A).
At each subsequent step, the algorithm places a single disk in a
position that maximally reduces the energy of the crystal interface. We
do not allow the particles to rearrange following placement. On a flat
surface, this algorithm produces a perfect crystal. Therefore any
deviation from a perfect crystal on a conical surface is the direct
result of geometric frustration, due to a $\delta$-function of Gaussian
curvature at the cone apex~\citep{zhang_fractional_2022}. Although this
algorithm does not account for the effects of temperature or kinetics,
it is a simple and effective way to model the effects of the closure
constraint on crystal growth. The details of the method are given in
Sec.~\ref{sec:methods}.

By modeling crystal growth in this manner, we find a proliferation of
defects (defined here as particles with anomalous coordination numbers)
for a crystal growing towards the tip of a cone with a small opening
angle, as shown in Sec.~\ref{sec:results}. As we shall show, this onset
of disorder results from a finite-size effect that depends strongly on
the local circumference and is insensitive to the seed orientation and
cone angle. Intuitively, the disordered regions appear when a
significant fraction of the growing interface consists of the seam of
the crystal. We develop a theoretical model that explains the results in
Sec.~\ref{sec:circumf} and discuss the influence of the crystal seed
location in Sec.~\ref{sec:escape} (the appendices provide additional
context, including discussions about corrections due to the
three-dimensional (3D) nature of the particles, commensurate packings on
cylinders, and a particularly interesting alternative seed composed of a
ring of particles). We conclude by noting that the transition to a
disordered packing can occur surprisingly far from the tip, which may
give some insights into the morphology of crystals seen in biology and
materials (Sec.~\ref{sec:conclusion}).

\section{Methods}
\label{sec:methods}

\begin{figure}
    \centering
    \includegraphics{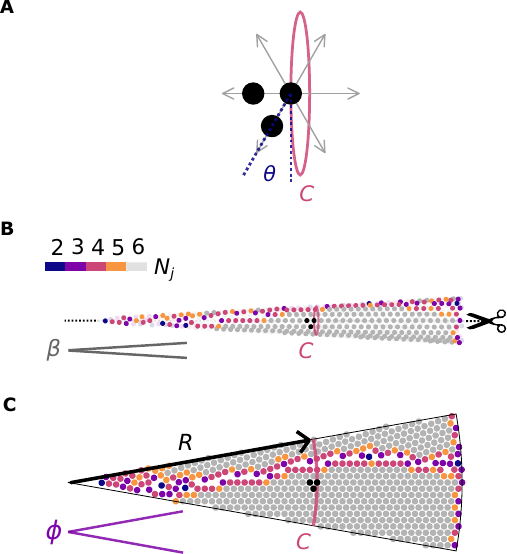}
    \caption{Geometric parameters in the simulation. (\textbf{A})
      Diagram of a triangular crystal seed. The angle $\theta$ describes
      the orientation of the lattice vector relative to the red cone
      circumference curve $C$. (\textbf{B}) Rendering showing results of
      a Bennett-type simulation~\citep{bennett_serially_1972} for a 2D
      packing of disks. We initialize the crystal seed (marked in black)
      on a 3D cone of angle $\beta$. The color of each particle
      indicates its coordination number $N_{j}$. (\textbf{C}) Rendering
      showing a mapping of the 3D cone in \textbf{B} to a 2D unrolled
      sector of angle $\phi=2\pi\sin(\beta/2)$, where the seed is at a
      sector radius $R$ from the apex. In this example,
      $\theta=30^{\circ}$, $\phi=20^{\circ}$, $C=12a_0$ and $R=54.4a_0$.
      A seam consisting of particles with coordination number $N_{j}<6$
      runs in the axial direction.}
    \label{fig:geometry}
\end{figure}

Our algorithm is designed to simulate an idealized crystallization
process in which a crystal grows from a single nucleus in a
reaction-limited fashion. To create the nucleus, we place three
close-packed particles with diameter $a_0$ in a tangent plane at a
radius $R=C/\phi$ from the sector vertex and with an orientation
$\theta$ with respect to the cone circumference $C$
(\cref{fig:geometry}A). We then add particles one by one, such that each
particle contacts the greatest number of other particles, or,
equivalently, forms the greatest number of ``bonds.'' A bond is formed
when the centers of two particles are less than $a_0$ apart with a
tolerance of $10^{-4} a_0$, representing an interaction potential with a
narrow attractive well, as is found in a number of colloidal
adsorption
experiments~\citep{meng_elastic_2014,tanjeem_geometrical_2021}. If there
are several degenerate options for placing the particle, we randomly
select one option. We do not place particles such that they form only
one bond with the existing assembly because the position of a dangling
bond is not well-defined, and a rotation of the dangling bond quickly
leads to contact with two particles. We also do not allow any previously
placed particles to move, nor do we let the entire crystal translate or
rotate.

We choose 2D circular disks of diameter $a_0$ to represent the effective
shapes of 3D spherical particles adsorbed onto a conical surface, and we
neglect the anisotropic, position-dependent stretching of particle
projections onto a conical surface. This approximation allows us to map
the three-dimensional (3D) cone of cone angle $\beta$ into a flat
two-dimensional (2D) circular sector with a periodic boundary condition
and a sector angle $\phi=2\pi\sin(\beta/2)$. The mapping is one-to-one
because the circumference $C$ of a circular cross-section of the 3D cone
(\cref{fig:geometry}B) is equivalent to the arc length $C$ of the 2D
sector at a given $R$ (\cref{fig:geometry}C). The resulting 2D algorithm
is computationally simpler, yet is still able to capture the effects of
the closure condition on frustration (see \cref{3Dcorrections} for a
discussion of the limitations of this approach).

We truncate the sector at a radial distance of $20a_0$ from the seed
position to encourage growth into the sector tip, where we expect to see
interesting structures. A simulation terminates when no more particles
can be placed into the sector or 1000 particles have been placed. For
comparison, the maximum number of particles that can pack into a
$\phi=5^{\circ}$ sector is around 800.

Since our aim is to understand the effects of geometrical frustration,
we select conditions under which perfect triangular crystals cannot
form. Perfect crystals lack seams and can form only when the sector
angle is a ``magic'' angle $\phi=60^{\circ} P$, where $P$ is an
integer~\citep{krishnan_graphitic_1997,
  ganser_barbie_k_assembly_1999,zhang_fractional_2022}. For example, a
cylinder is a $P=0$ magic cone with $\phi=0^\circ$. We therefore
restrict our study to cones with $\phi\neq60^{\circ}P$. A seam on such a
cone is shown in \cref{fig:geometry}B and~C. Particles at the crystal
self-boundary have coordination number $N_{j}<6$. For particles with
Lennard-Jones-like pair potentials where the range of attractive
interaction is comparable to the defect hard-core diameter, one might
expect a grain boundary to form~\citep{bausch_grain_2003}. Our
simulations, however, have an interaction range of order $10^{-4}$ times
the hard-core diameter, and the resulting seam, like a stacking fault,
maintains its integrity.

We choose to study near-cylindrical cones with small opening angles of
$\phi\leq30^\circ$ to facilitate comparison with results of previous
studies of crystals on cylinders~\citep{wood_self-assembly_2013,
  tanjeem_geometrical_2021}. When simulating growth on a surface with
$\phi=0^\circ$, corresponding to a true cylinder, we select a seed
orientation $\theta$ such that seams are still geometrically required.

To characterize the structures, we calculate the bond orientational
order parameter $\psi_{6,j}$ for each particle $j$ with nearest
neighbors indexed by $k$~\citep{han_melting_2008, halperin_theory_1978}:
\begin{equation}
\psi_{6,j} =
\frac{1}{N_{j}}\sum_{k=1}^{N_{j}}e^{i6\theta_{jk}},
\end{equation}
where $\theta_{jk}$ is the angle between the circumferential axis and
the vector from particle $j$ to nearest neighbor $k$, and $N_{j}$ is the
number of nearest neighbors, or coordination number, of particle $j$. We
consider only particles that are separated by $a_0 \pm 10^{-4} a_0$ as
nearest neighbors.

We also calculate the defect density $\rho$ as a function of distance
$R$ from the vertex. We define defects as particles that have
$\left|\psi_{6,j}\right|<0.9$. This definition allows us to distinguish
defects, which disrupt the order of the crystal, from particles on the
boundary of the seam, which are part of an ordered crystal. Since the
same defect trends are preserved for different cutoff values of
$\left|\psi_{6,j}\right|$, we choose a high cutoff value to obtain a
sensitive measure of the defect density (see \cref{sec:defect-def}). To
calculate the defect density, we first bin the particles by $R$. We then
calculate the number of defects per number of particles within each bin
of width $a_0$, averaged over 100 trials.

\section{Results}
\label{sec:results}
\begin{figure}
    \centering
    \includegraphics{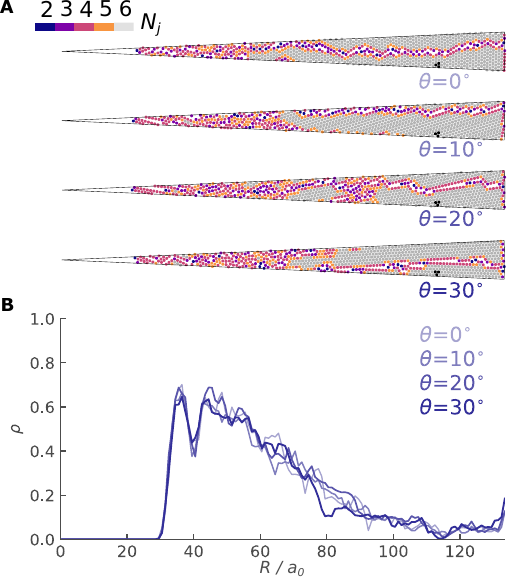}
    \caption{Variation of particle packings with seed orientation
      $\theta$ on cones with sector angle $\phi$. (\textbf{A}) Rendering
      of simulation results at unrolled cone angle $\phi=5^{\circ}$ show
      that particles in the vicinity of the seed placed at $R=114.59a_0$
      $(C=10a_0)$ are ordered. As the crystal approaches the vertex, the
      packing becomes disordered. (\textbf{B}) Plot of the defect
      density $\rho$ as a function of radial distance $R$ from the tip,
      where $\rho$ is the number fraction of particles with
      $\left|\psi_{6,j}\right|<0.9$ in a given $R$ bin averaged over 100
      trials at each seed orientation $\theta$. The defect density
      curves are similar for all $\theta$ at $\phi=5^{\circ}$. Note that
      the defect density rises from the seed toward the tip and then
      drops to zero at $R\approx28.5a_0$ because the particles are
      squeezed out of the tip.}
    \label{fig:varytheta}
\end{figure}

We first explore the effects of seed orientation on the structure of the
system. We find that the crystal initially grows from the seed as an
ordered packing of particles, represented by grey particles with
$N_{j}=6$ in \cref{fig:varytheta}A. As the crystal wraps around the
cone, it meets itself to form a seam consisting of particles with
$N_{j}<6$. However, as the seam approaches smaller circumferences, these
defects begin to dominate the growth interface, leading to the formation
of a disordered region consisting primarily of defect particles with
$N_{j}<6$ (see \cref{fig:varytheta}A, regions near tips of cones).

For all seed orientations, the defect density follows a similar curve
(\cref{fig:varytheta}B), tending to increase as $R$ decreases. The local
minimum at approximately $R=40a_0$ corresponds to the location where a
crystalline cluster three particles wide can form. The defect density
then increases again for smaller $R$, until it reaches a maximum value
and falls rapidly to $\rho=0$ at approximately $R=28.5a_0$. Note that
$R\approx23a_0$ represents the limit for packing two particles side by
side on unrolled cones of angle $\phi=5^{\circ}$. Overall, the tendency
of the defect density to increase as the cone narrows suggests that
finite-size effects are responsible for the disorder near the tip.

\begin{figure}
    \centering
    \includegraphics{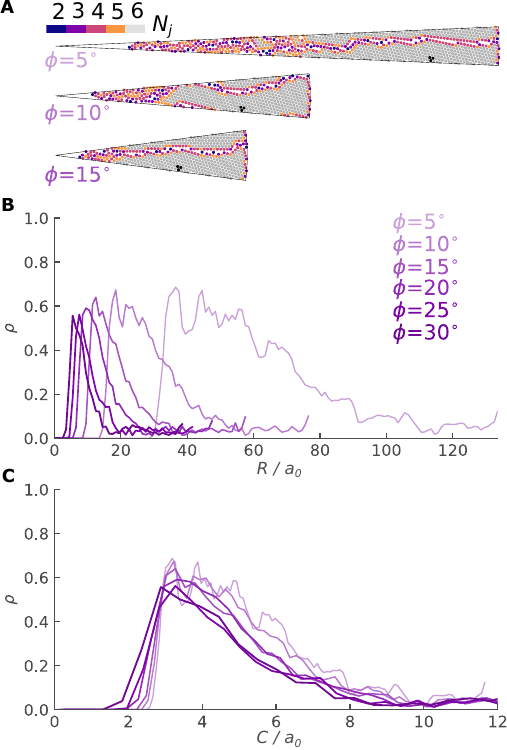}
    \caption{Variation of particle packings with unrolled sector angle
      $\phi$ at fixed seed orientation $\theta$. (\textbf{A}) Renderings
      of simulation results show a long region of $N_j<6$ particles at
      $\phi=5^{\circ}$, while the disordered regions are concentrated
      closer to the tip at $\phi=10^{\circ}$ and $\phi=15^{\circ}$. The
      seed crystals are at $C=10a_0$ with $\theta=40^{\circ}$.
      (\textbf{B}) Plot of the defect density $\rho$ as a function of
      $R$, averaged over 100 trials. The distribution is broad for
      $\phi=5^{\circ}$ and becomes narrower for increasing sector
      angles. (\textbf{C}) For each $\phi$, $\rho$ is mapped to $C$ by
      $C=R\phi$. $\rho$ collapses as a function of $C$. }
    \label{fig:varyphi}
\end{figure}

When we initialize crystals on cones with different sector angles with
fixed seed orientation $\theta=40^{\circ}$, we find that the position at
which the disordered region emerges also varies, as seen for unrolled
cones in \cref{fig:varyphi}A. For small sector angles, we find more
disorder farther from the tip, while for larger sector angles, the
disordered region is found nearer the tip (\cref{fig:varyphi}B). These
results show that cones with small angles can have long disordered
regions.

By rescaling $R$ by $R\phi=C$, we find that the defect density collapses
as a function of cone circumference (\cref{fig:varyphi}C). Defect
proliferation therefore depends on the circumference and not strongly on
the sector angle or seed orientation. Although the precise sector angle
and seed orientation might alter the details of the closure constraint
at the single-particle scale, the circumference at which closure occurs
has a greater effect on crystal growth.

The density curves do not perfectly collapse because of our discrete
binning procedure, which results in a systematic variation with the
sector angle. As $\phi$ increases, the gradient of C also increases. An
annular bin of fixed width centered at C will not only access larger
circumferences but also have more particles at the larger circumference,
which are less likely to be defects. Therefore, at a given C, the defect
density for the annular bin is biased towards a lower $\rho$ as $\phi$
increases. It is possible that the gradient of $C$ could also affect the
defect density in other ways. Nonetheless, the near-collapse of the
density curves upon rescaling shows that the circumference, rather than
the gradient, is the most important parameter to consider.

\begin{figure}
    \centering
    \includegraphics{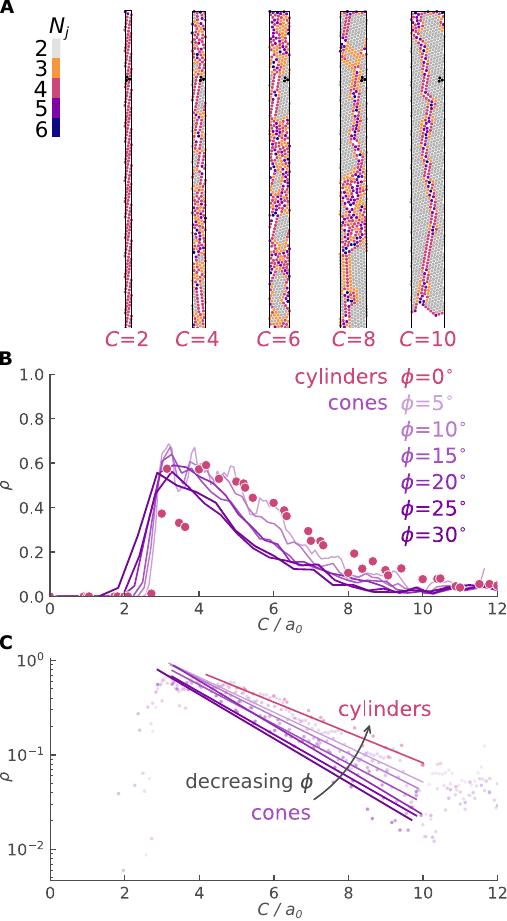}
    \caption{Particle packings on 2D cylinders of different
      circumferences, measured in units of the hard core diameter $a_0$.
      (\textbf{A}) Rendering of representative results for simulations.
      At $C=10a_0$, the crystal consists of primarily $N_j=6$ particles.
      As $C$ decreases, the crystal becomes fragmented and disordered.
      The seed orientation is fixed at $\theta=40^{\circ}$. (\textbf{B})
      Plot of defect density of cylinders (circles) superimposed on the
      plot for cones from \cref{fig:varyphi}C (lines). The cylinder
      defect densities are calculated as the number fraction of defects
      relative to the total number of particles, averaged over 100
      trials at each circumference. (\textbf{C}) As the circumference
      decreases, the defect density grows exponentially at small
      circumferences for both cones and cylinders. The chosen
      circumferential values $C/a_0=$2, 4, 6, 8, and 10 in this figure
      do not include any of the special values that would result in
      perfect packings for $\theta=40^{\circ}$
      (\cref{sec:magic})~\citep{beller_plastic_2016}.}
    \label{fig:cylinders}
\end{figure}

To show that defect proliferation can be understood as a function of
circumference, we examine crystal growth on cylinders, which have a
constant circumference. We find that cylinders with large circumferences
have lower defect densities than cylinders with small circumferences
(\cref{fig:cylinders}A). For large cylinders, the crystal forms a seam,
as expected, and is ordered with few defects. As the circumference
decreases, however, the crystal becomes increasingly fragmented as more
defects are incorporated into the packing. For thin cylinders, the
packing is predominantly disordered.

We find that the defect densities on cylinders as a function of
circumference follow the same trend as the defect densities of cones
(\cref{fig:cylinders}B), provided the orientation $\theta$ of the
triangular seed cluster is not tuned to the special phyllotactic value
that allows a commensurate tiling by a triangular
crystal~\citep{harris_tubular_1980, mughal_theory_2014,
  beller_plastic_2016}. Because our focus here is on seeds with random
orientations, we neglect the interesting regular tilings that occur for
commensurate seed orientations at fixed cylinder radius, or commensurate
cylinder radius at fixed seed orientation (\cref{sec:magic}). For seed
orientations $\theta=40^{\circ}$, we find that the defect density
distributions appear exponential for small $R$, regardless of whether
the surfaces are conical or cylindrical (\cref{fig:cylinders}C).
Crystals on cylinders therefore reproduce the finite-size effects seen
in crystals on cones, if special commensurate tilings are
ignored~\citep{harris_tubular_1980, mughal_theory_2014}.

\section{Discussion}
\label{sec:discussion}
\subsection{Circumference determines defect density}\label{sec:circumf}
To provide some intuition for exponential defect proliferation observed
in our simulations at small cone circumferences, we introduce a simple
theoretical model for slow, reaction-limited growth on a
\emph{cylindrical} substrate. We expect our algorithm to simulate slow,
reaction-limited growth because particles attach one-by-one to the
growing crystal, and each added particle minimizes the local energy of
the crystal interface. We use a cylindrical substrate to simplify our
theoretical arguments because, as shown above, cylinders reproduce the
finite-size effects seen on cones.

\begin{figure}
\begin{center}
\includegraphics{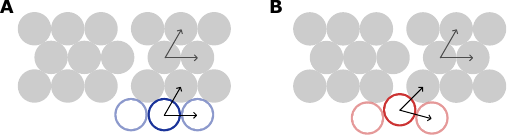}
\caption{Illustration of the defect growth process at a seam. Particles
  in the preexisting crystal are shown as filled gray circles. When all
  candidate sites result in the formation of two bonds, a particle can
  attach randomly at (\textbf{A}) a ledge site that initiates a new
  crystal row or (\textbf{B}) a seam site that creates more sites that
  break the symmetry of the preexisting crystal. Dark open circles show
  new particle locations, with lighter open circles emphasizing
  candidate sites created by attachment of the new particle. Note that
  this schematic depicts growth on a cylinder---the crystal rows on
  either side of the seam are parallel.}
\label{fig:defectrow}
\end{center}
\end{figure}

Given a crystal with a seam and smooth facets as in
\cref{fig:defectrow}, we consider the types of lattice sites that an
additional particle can diffuse to in the context of a 2D
terrace-ledge-kink model, used to describe ideal surface crystal
growth~\citep{nozawa_step_2018}. Particles sitting at the crystalline
edges form the ledges where new particles can adsorb. Kinks describe
missing particles along the ledge. In our simulation of slow ideal
growth, kinks, which have three or more dangling bonds by definition,
are higher-energy sites than the smooth ledges, which have two dangling
bonds. Therefore, a particle diffusing to the crystal will adsorb to
kinks first.

Once the kink sites have been filled, a particle can attach to two types
of energetically equivalent lattice sites. The first, a ledge site, is
continuous with the preexisting crystal (\cref{fig:defectrow}A, dark
blue circle). The second, a seam site, is incompatible with the
preexisting crystal (\cref{fig:defectrow}B, dark red circle).

Particle attachment to a ledge site results in on-lattice growth,
meaning that the symmetry of the preexisting crystal is preserved.
Particle attachment to a seam site results in off-lattice growth,
meaning that the symmetry of the preexisting crystal is broken. We
expect the density of disordered defects to be related to the proportion
of seam sites to ledge sites. Crucially, off-lattice growth increases
the probability of further off-lattice growth because adsorption of a
seam particle increases the number of off-lattice candidate states
overall (light red circles in \cref{fig:defectrow}). Thus, if
off-lattice growth is likely, this argument predicts that it will only
become more likely as growth proceeds, initiating the formation of a
disordered region as we see in our simulations.

To further develop this simplified picture, we estimate the probability
of disordered growth. A perfect crystal with rows composed of
$N_\textnormal{row}$ particles and a single seam as in
\cref{fig:defectrow} has of order $N_\textnormal{row}$ candidate sites
that lead to on-lattice growth, and only order one candidate sites that
lead to off-lattice growth. Therefore, the probability of on-lattice
growth occurring initially is $P_1(t=0) \sim (1-1/N_\textnormal{row})$,
and the probability of off-lattice growth is $P_2(t=0)\sim
1/N_\textnormal{row}$. If on-lattice growth occurs, $P_1$ and $P_2$ do
not change. If off-lattice growth occurs, we expect that $P_2$ increases
by an amount that scales with $1/N_\textnormal{row}$. If $P_2$ reaches
some threshold value---say $P_2 = 1/2$, at which half of the candidate
sites lead to off-lattice growth---runaway off-lattice growth results,
leading to the formation of a disordered region.

What is the probability of $P_2(t)$ increasing to this threshold value?
If on-lattice growth occurs following off-lattice growth, there may be
some healing of the disordered seam region, and $P_2$ may decrease. We
therefore make the simplifying approximation that $P_2$ increases only
when off-lattice growth occurs many times in a row, and $P_2$ increases
by $1/N_\textnormal{row}$ every time off-lattice growth occurs
consecutively. With these assumptions, the probability of off-lattice
growth occurring $s$ consecutive times scales as
\begin{equation}
\begin{aligned}
    P_s={} & P_2(t=0)\,P_2(t=1)\ldots P_2(t=s) \\ 
           & \sim \frac{1}{N_\textnormal{row}}\times \frac{2}{N_\textnormal{row}} \times \ldots \times  \frac{s}{N_\textnormal{row}}=\frac{s!}{\left(N_\textnormal{row}\right)^s}.
\end{aligned}
\end{equation}
To find the probability that $P_2$ reaches $1/2$, we let
$s=N_\textnormal{row}/2$ and make Stirling's approximation:
\begin{equation}
    P_{N_\textnormal{row}/2}= \frac{(N_\textnormal{row}/2)!}{\left(N_\textnormal{row}\right)^{N_\textnormal{row}/2}} \sim  \frac{\sqrt{N_\textnormal{row}}}{2^{N_\textnormal{row}/2}}e^{-N_\textnormal{row}/2}.
\end{equation}
Therefore, this simple model predicts that the probability of a
disordered region initiating increases exponentially as the number of
particles in a crystal row encircling the cylinder decreases. The
predicted exponential scaling is consistent with the simulation results,
which show that the defect density increases exponentially with
decreasing circumference (\cref{fig:cylinders}C). The same argument can
explain defect formation on a cone, albeit with some subtleties, which
we discuss in \cref{sec:circumf-cone}.

\subsection{Seeds close to the tip}\label{sec:escape}

\begin{figure}
    \centering
    \includegraphics{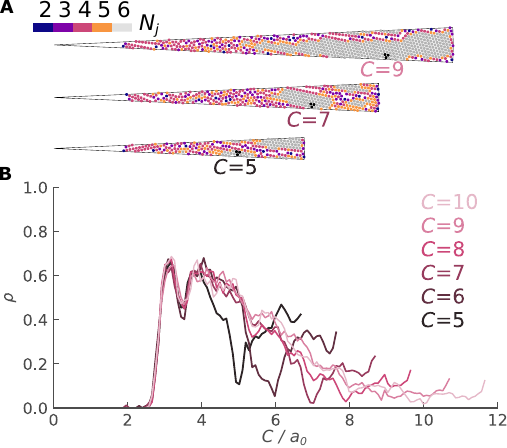}
    \caption{Particle packings on sectors initialized at different seed
      positions. (\textbf{A}) Rendering of simulation results for
      $\phi=5^{\circ}$ with seed positions at $C=5a_0$, $C=7a_0$, and
      $C=9a_0$ and a seed orientation of $\theta=40^{\circ}$.
      (\textbf{B}) Plot of the defect density $\rho$ as a function of
      $C$, averaged over 100 trials. Crystals seeded at small
      circumferences have dips in the defect density that correspond to
      the seed location.}
    \label{fig:varycinit}
\end{figure}

These results, all of which concern seeds placed far from the tip of the
cone, raise the question of whether placing seeds closer to the tip of
the cone might help prevent disorder near the tip. We therefore examine
simulations of crystallization with seeds placed at different
circumferences.

We find that while crystals seeded far from the tip (at $C=9a_0$) are
able to grow normally towards the tip until the onset of disorder,
crystals that are seeded closer to the tip (at $C=7a_0$ and $C=5a_0$)
first form a small crystal before disorder emerges
(\cref{fig:varycinit}A). The formation of these small crystals is
reflected in the defect density, which shows a dip at the circumference
corresponding to the seed location for all seeds placed at $C<9a_0$
(\cref{fig:varycinit}B).

We can explain these results using the model from \cref{sec:circumf}. A
crystal grows until the closure constraint demands formation of a seam.
But when the crystal is seeded near the tip, the circumference is small,
and hence $N_\textnormal{row}$ is small. Therefore off-lattice growth is
more probable at these small circumferences, and the crystal becomes
frustrated a short distance from the seed.

Interestingly, in some cases we see that new crystals can form at the
wider part of the cone, as shown in the $C=7a_0$ example. However, these
new crystals quickly become frustrated again as new seams and grain
boundaries form. Consequently, the defect density dips at the
circumference of the seed and then rises with increasing circumference,
until it exceeds the defect density for a seed placed at $C=10a_0$
(\cref{fig:varycinit}B). We conclude that crystals that are seeded near
the tip can temporarily escape the finite-size effect leading to
disorder near the tip, though at the expense of increased disorder
farther from the seed site.

\section{Conclusion}\label{sec:conclusion}

We have shown that crystal growth on a cone is geometrically frustrated.
For any non-magic cone angle, a seam is required. A disordered region
forms near the tip because defects tend to appear at the seam, and the
probability of these defects proliferating increases exponentially as
the circumference decreases.

This type of frustration has implications for slow, reaction-limited
crystal growth on cones. Near-cylindrical cones have long sections in
which defects form with high probability, resulting in large areas of
potentially disrupted crystallization. In wider cones, the increase in
defect probability is concentrated at the tip and can block tip closure,
leading to holes at the tips of conical shells.

These results may help explain tip-closure problems observed in
experimental systems. For example, the conical capsids of HIV often
exhibit large holes at the tip~\citep{ganser_barbie_k_assembly_1999}.
Also, crystals of WS$_2$ have been found to terminate unexpectedly far
from the tip~\citep{yu_tilt_2017}. Future experiments on colloidal
systems, such as the system described in Chapter~6 of
Ref.~\citenum{tanjeem_effect_2020}, might shed light on whether
tip-closure failures are the result of the frustration mechanism
revealed by our simulations.

Our results also show that control over nucleation may be crucial to
fabricating conical crystals for applications. Disordered regions form
near the tip of the crystal, regardless of whether the seed is far or
close to the tip. But a crystal seeded near the tip can temporarily
bypass the finite-size effect, resulting in a locally reduced defect
density. Therefore, if the surface or interactions can be controlled
such that nucleation occurs close to the tip, at least small crystals
can be formed in this region. Furthermore, if the geometry of nucleation
can be controlled, new crystalline structures might be realized
experimentally. In \cref{sec:nucleation}, we discuss how a nucleus
consisting of a ring of particles might grow, and how the size of the
resulting crystal depends on the elastic modulus.

Our simulations used a greedy algorithm because our aim was to reveal
the geometric frustration faced by crystals growing on a conical
surface. Our simulations do not account for kinetics, thermal
fluctuations, or vibrational entropy. Future simulations and experiments
are therefore needed to develop a more complete physical understanding
of conical crystal growth. Nonetheless, our results show that, apart
from the special case of magic-angle cones, any conical crystal is
subject to geometrical frustration that promotes disorder at small
circumferences.

\section*{Data Availability Statement}
Data for the simulations are openly available on the Harvard
Dataverse~\cite{sun_data}. Code for the simulations is available under
the GNU General Public License v3 at
\url{https://github.com/manoharan-lab/cone-disk-packings}.

\begin{acknowledgments}
  We thank Lara Braverman for insightful conversations. This research
  was primarily supported by the National Science Foundation through the
  Harvard University Materials Research Science and Engineering Center
  under grant number DMR-2011754. Additional support was provided by the
  National Science Foundation Graduate Research Fellowship Program under
  grant numbers DGE-2140743 and DGE-1745303.
\end{acknowledgments}

\clearpage
\def\bibsection{\section*{\refname}}
%

\clearpage
\setcounter{figure}{0}
\renewcommand{\figurename}{Fig.}
\renewcommand{\thefigure}{S\arabic{figure}}

\appendix
\section{3D corrections to effective particle shape}\label{3Dcorrections}
Using a constant particle shape simplifies our analysis considerably.
However, a position-dependent particle shape would more accurately model
3D spheres assembling onto a cone. Though we do not consider these
corrections in this work, we describe them briefly here.

\begin{figure}
\begin{centering}
\includegraphics[width=0.8\columnwidth]{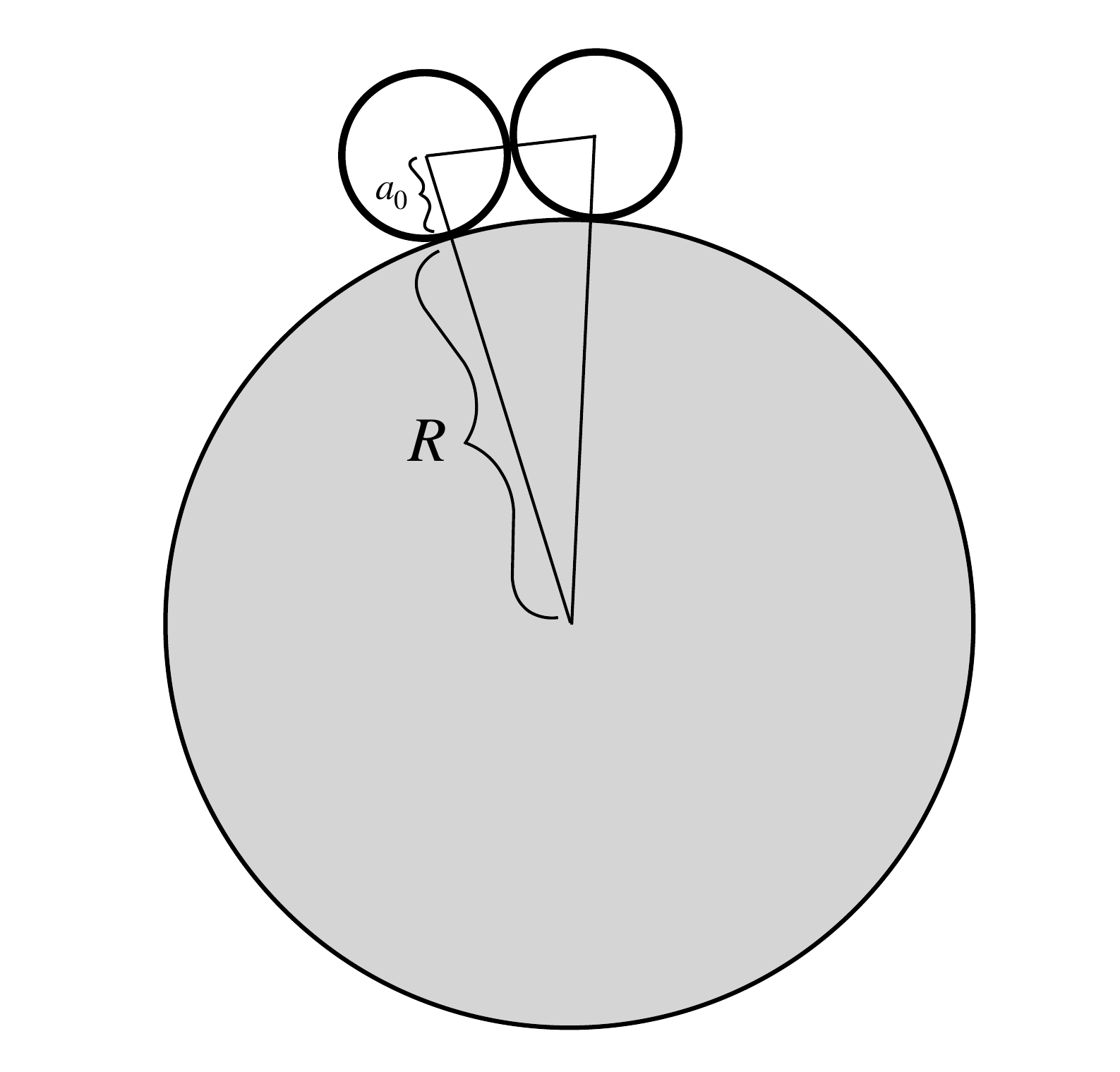}
\caption{Diagram showing that the distance between particles along a
  surface differs from $2 a_0$ when the surface is curved.}
\label{fig:circlepack}
\end{centering}
\end{figure}

To provide intuition, we first consider two circular particles tangent
to each other and to the surface of a circle, as shown in
\cref{fig:circlepack}.
The distance between particle centers is $d_\textnormal{ctr}=2 a_0$.
However, measured along the surface, the distance between the two points
at which the particles touch the substrate is
\begin{equation}
  \begin{aligned}
    d_\textnormal{surf} &=2 R \sin^{-1}\left(\frac{a_0}{R+a_0} \right) \\
  &\approx 2 a_0 - \frac{2 a_0^2}{R} + \mathcal{O}\left(\frac{a_0^3}{R^2}  \right).
\end{aligned}
\end{equation}
In the limit $a_0/R \to 0$, $d_\textnormal{ctr}= d_\textnormal{surf}$,
while for finite $R$, $d_\textnormal{surf}<d_\textnormal{ctr}.$ This
difference can be significant: for $R=3 a_0$, $d_\textnormal{surf}$ is
approximately 75\% of $d_\textnormal{ctr}$, for example. We can represent
this 2D assembly in 1D by ``unwrapping'' the circle to form a line
segment of length $2 \pi R$ with periodic boundaries, upon which we
place two adjacent line segments of length $d_\textnormal{surf}$. The
dimensional reduction creates an effective particle size that depends on
the surface curvature.

This idea can be straightforwardly extended to a cylindrical substrate
with radius $R$, as described in~\citet{memet_random_2019}. Two adjacent
particles aligned with the axial direction are separated by a distance
$d_\textnormal{ctr}$, while two adjacent particles aligned with the
circumferential direction are separated by a distance
$d_\textnormal{surf}$ measured along the surface. Upon unwrapping the
cylinder to form a flat 2D domain with periodic boundaries, we find that
the effective particle shape is an extended, ellipse-like shape. When
centered at the origin of the $xz$-plane, this effective shape is
described by the equation
\begin{equation}
    (a_0+R)^2 \sin^2(x/R) +z^2 = a_0^2.
    \label{eq:oblong}
\end{equation}
Note that this is the equation of a circle in the limit $R \gg a_0$, as
expected. Thus, we can simulate crystallization on a cylinder using a
flat 2D domain while taking into account the 3D shape of the particles
if we generalize our algorithm to use the oblong shapes given by
\cref{eq:oblong} instead of circles.

For a cone, the effective particle shape is more complicated. Far from
the cone tip, curvature effects will be small, and the effective
particle shape will be approximately circular. As we move closer to the
tip, curvature along the circumferential direction increases, and the
effective particle shape will become more elongated.

\begin{figure}
\begin{centering}
\includegraphics[width=0.8\columnwidth]{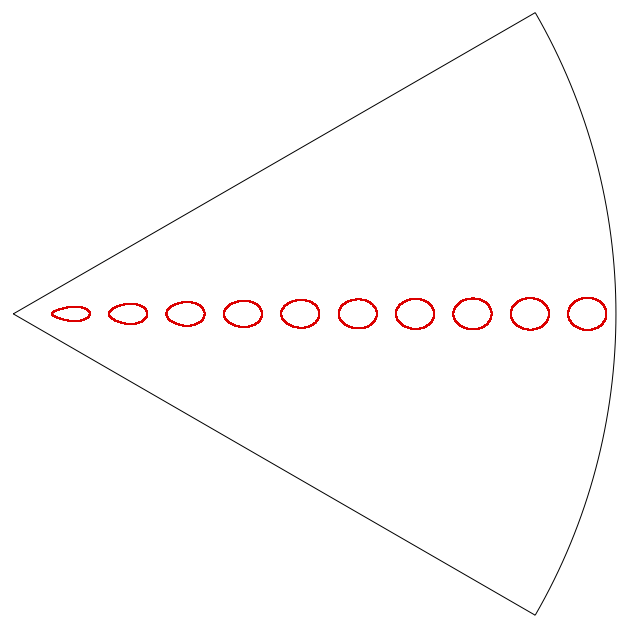}
\caption{Diagram showing that the effective particle shape on a cone is
  a tapered ellipse-like shape, pointing towards the sector center. The
  correction is much more pronounced close to the tip.}
\label{fig:wedges}
\end{centering}
\end{figure}

We can solve for this position-dependent effective particle shape by
considering horizontal slices through the cone and a tangent spherical
particle. Effective particle shapes for different locations on the
domain are shown in \cref{fig:wedges}.

Using a position-dependent effective particle shape will likely further
disrupt crystallization close to the sector center/apex (where the
particle shape is not approximately circular). These 3D effects should
be considered by simulations that seek to more accurately model
experiments. Though solving explicitly for the effective particle shape
in \cref{fig:wedges} is useful for estimating the corrections to our
model, simulating the assembly process directly in three dimensions
would be a more straightforward way to incorporate this effect.

\section{Defect definition}\label{sec:defect-def}
\begin{figure}
\begin{centering}
\includegraphics{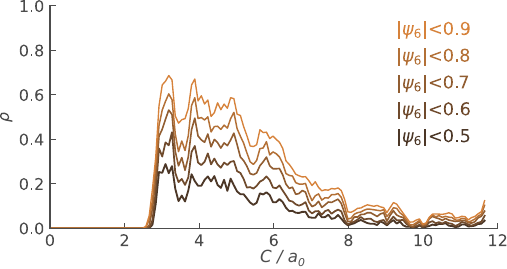}
\caption{Plots of defect density for various choices of cutoff for
  $\left|\psi_{6,j}\right|$. The simulations were carried out for
  sectors with $\phi=5^{\circ}$ and for seeds initiated at $C=10 a_0$
  with orientation $\theta=40^{ \circ}$. Each curve is an average over
  100 trials.}
\label{fig:psi6cutoff}
\end{centering}
\end{figure}

We choose $\left|\psi_{6,j}\right|<0.9$ to define a defect. We use the
bond orientational order instead of the more commonly used definition
based on coordination number so that we can filter out dislocations that
form along the seam, which have $\left|\psi_{6,j}\right|\geq0.9$. These
dislocations would otherwise contribute a background defect density.
Changing the cutoff value from 0.9 does not alter the shape of the
defect density distribution, as seen in \cref{fig:psi6cutoff}.

\section{Commensurate packings on cylinders}\label{sec:magic}
\begin{figure}
    \centering
    \includegraphics{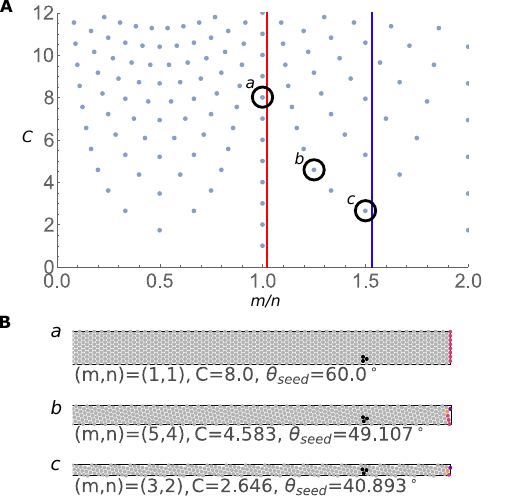}
    \caption{Commensurate chiral and achiral particle packings on
      cylinders. (\textbf{A}) Plot of Eq.~\ref{eq:commpoints} which
      represents the circumferences $C$ and seed orientations $\theta$
      that should result in commensurate packings, where $\theta$ is
      related to $(m,n)$ by Eq.~\ref{eq:thetamn}. The blue vertical
      line represents a seed orientation of $\theta=40^\circ$ as shown
      in \cref{fig:cylinders} of the main text, while the red vertical
      line represents a seed orientation of $\theta=59^\circ$, as shown
      in \cref{fig:nonmono}. Note that the vertical lines do not
      intersect any commensurate points for $C<20a_0$. Here we show
      commensurate points for $C<12a_0$, which is the circumference range
      studied. (\textbf{B}) Renderings of simulation results on an
      unrolled cylinder with crystal seed parameters set at commensurate
      points. We observe perfect packings even for chiral seed
      orientations and small circumferences.}
    \label{fig:magic}
\end{figure}
\begin{figure}
    \centering
    \includegraphics{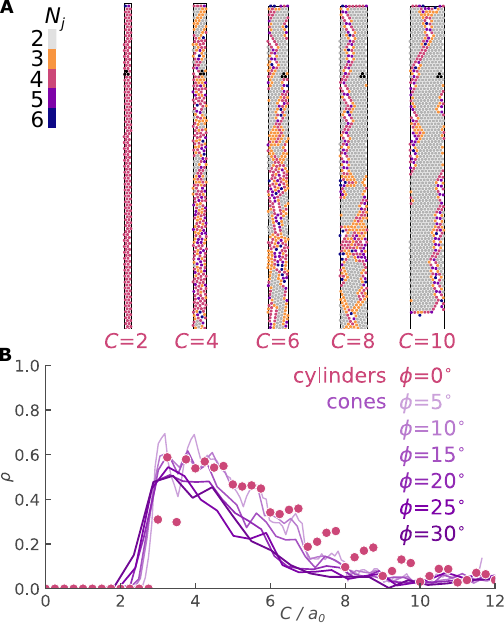}
    \caption{Nonmonotonic defect density behavior of near-commensurate
      cylinders. (\textbf{A}) Rendering of representative results for
      simulations on an unrolled cylinder for seed orientation fixed at
      $\theta=59^\circ$. As $C$ decreases, the crystal becomes
      fragmented and disordered. (\textbf{B}) Plot of the defect density
      of cylinders (circles) superimposed on a plot for cones (lines)
      for simulations fixed at $\theta=59^\circ$. Note that the cylinder
      defect densities are nonmonotonic, with local minima at integer
      $C$ values.}
    \label{fig:nonmono}
\end{figure}

On a cone, perfect crystals can form only when the sector angle is a
``magic'' angle $\phi=60^{\circ} P$, where $P$ is an
integer~\citep{krishnan_graphitic_1997, ganser_barbie_k_assembly_1999,
  zhang_fractional_2022}. The only magic angle in the range
$\phi\leq30^\circ$ studied here is $\phi=0^\circ$, corresponding to a
cylinder. On a cylinder, there are many commensurate packings, depending
on the circumference and angle of the crystal. These commensurate
packings are determined by a pair of phyllotactic indices $(m,n)$
where~\citep{beller_plastic_2016}
\begin{equation}
    \frac{1}{2\pi} \left| C \right|=\frac{a_0}{2\pi}\sqrt{m^2+n^2-mn}
    \label{eq:commpoints},
\end{equation}
and $a_0$ is the lattice spacing. The integers $m$ and $n$ are related
to the crystal orientation by~\citep{beller_plastic_2016}
\begin{equation}
    \tan{\left(\frac{\pi}{2}-\theta\right)}\approx\frac{2}{\sqrt{3}}\left(\frac{m}{n}-\frac{1}{2}\right)
    \label{eq:thetamn},
\end{equation}
where $\theta$ is the angle between the crystal lattice and
circumferential axis.

The commensurate points in the $(C,m/n)$ plane are plotted in
\cref{fig:magic}A according to Eq.~\ref{eq:commpoints}. The blue
vertical line represents the crystal orientation $\theta=40^\circ$.
Along this line there are no commensurate packings, at least for $C<20a_0$. The red vertical line represents crystal orientations
$\theta=59^\circ$. Along this line there are also no commensurate
packings for $C<20a_0$. The circled points are visualized in
\cref{fig:magic}B. Our simulations are able to reproduce the
commensurate packings at the expected phyllotactic values.

When the seed orientation is fixed at a near-commensurate value of
$\theta=59^\circ$, our simulations reveal interesting non-monotonic
behavior in the defect density plot (\cref{fig:nonmono}). While the
overall boundary of the sampled defect densities follows the same
profile as cones, we observe local dips at integer $C$ values,
corresponding to values close to commensurate points
(\cref{fig:magic}A). Therefore, while cylinders of different crystal
orientations manage to capture the general defect density behavior of
cones, we choose to analyze seed orientations $\theta=40^\circ$, for
which we encounter fewer nearby commensurate points as cylinder
circumference increases (see blue line in \cref{fig:magic}).

\section{Circumference argument applied to cones}\label{sec:circumf-cone}
We now seek to extend the growth probability argument presented in
\cref{sec:circumf} to a conical substrate.
On a cone, crystals grow along conical geodesics. The theory for
cylinders might need to be modified if the number of particles along a
geodesic interface were to differ significantly from the number of
particles that can fit around the cone circumference at a given height,
or if the interface/geodesic length were to depend strongly on the
crystal orientation.

\begin{figure}
\begin{centering}
\includegraphics[width=\columnwidth]{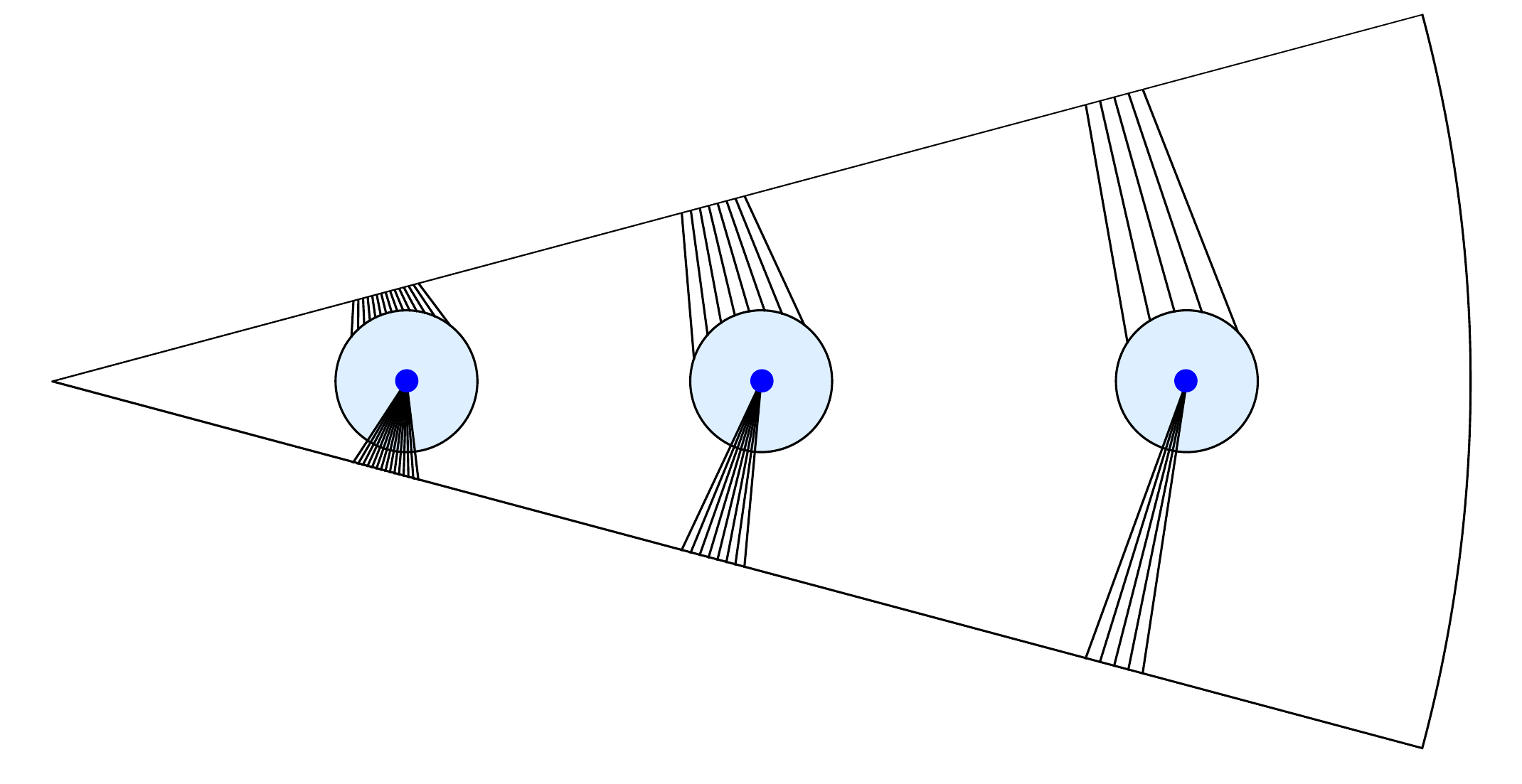}
\caption{Diagram showing examples of geodesics on the circular sector
  that make permissible crystal boundaries. Three possible starting
  points are shown (blue points) from which geodesics depart toward the
  bottom at a range of angles, circle the cone once, and return within a
  distance $a_0$ of the initial starting point (light blue disk). The
  lengths of the geodesics that meet these conditions are
  well-approximated by the circumference of a circular cross-section of
  the cone at the initial points.}
\label{fig:geodesics}
\end{centering}
\end{figure}

However, we find that if we restrict our attention to geodesics that
represent realistic crystal boundaries (for example, the geodesics that
start at a particular point on the cone, circle the cone once, and
returns to within $a_0$ of the starting point) on cones with small
angles, the length of a geodesic is well-approximated by the
circumference of a circular cross-section. The simplest case is a
geodesic at an angle $\theta=0$. The geodesic length is $2 R
\sin(\phi/2)$, which is the same as the circumference at that point, $R
\phi$, in the small-angle limit. Perhaps more surprisingly, varying
$\theta$ leads to only a minor correction. Only a narrow range of
$\theta$ values produce curves that circle the apex one time, returning
to within $a_0$ of their starting point, and these admissible curves are
also well-approximated by the circumference, as we show graphically in
\cref{fig:geodesics}. It is therefore a reasonable approximation to set
$N_\textnormal{row} \approx \text{circumference}/a_0$ and to explain the
dependence on circumference (and independence on seed orientation and
cone angle) in our cone simulations with our simplified growth model
developed for cylinders.

\section{Initialization with a ring of particles} \label{sec:nucleation}

\begin{figure}
\begin{centering}
\includegraphics[width=0.5\columnwidth]{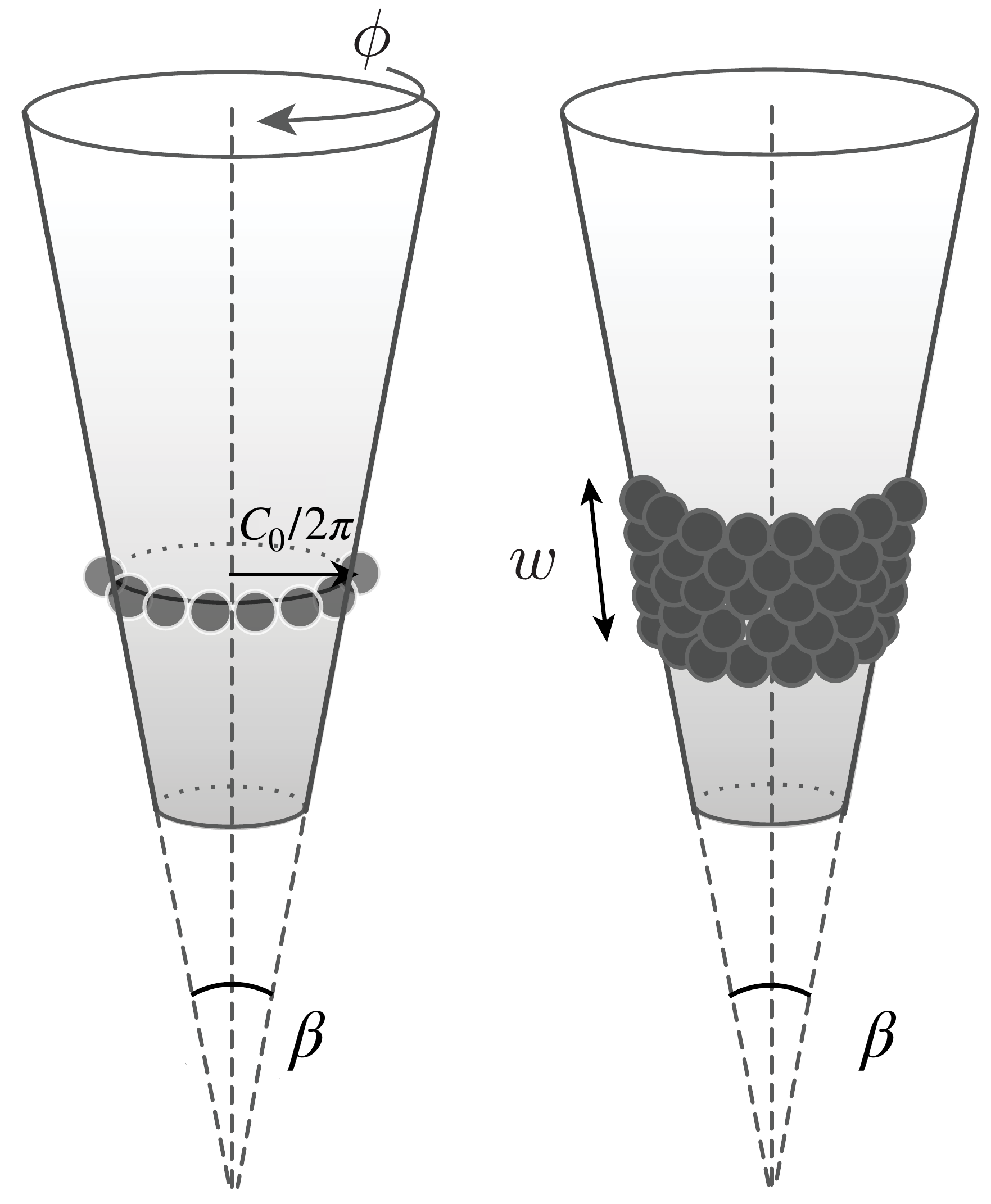}
\caption{Schematic of ring initial condition, defining the ring
  circumference $C_0$ and the banded crystal width $w$.}
\label{fig:ring}
\end{centering}
\end{figure}

In the main text, we investigate crystal growth initiated by a seed of
three particles in contact with one another (Fig.~\ref{fig:geometry}A).
In this appendix, we provide a brief overview of the interesting effects
that occur when a crystal instead nucleates on a ring of particles
oriented parallel to the base of the cone (Fig.~\ref{fig:ring}).
Such a ring-shaped crystal could be realized experimentally by, for
example, milling a shallow trench or placing a metallic ring flush
against the conical surface. Reliably creating ring-shaped initial
conditions in experiments is the subject of ongoing work.

First, we consider a ring of particles initiating crystal growth on a
cylindrical substrate. A crystalline band of any width is not strained
on a cylinder provided that the initial seeded ring is commensurate with
the circumference of the cylinder. The energy of a crystal of
circumference $C_0$ and width $w$ is given by
\begin{equation} \label{eq:Ecylinder}
    E_\mathrm{cylinder} = E_\mathrm{bound} + E_\mathrm{bulk},
\end{equation}
where
\begin{eqnarray}
    E_\mathrm{bound} &=& 2 \sigma C_0 \\
    E_\mathrm{bulk} &=& -  \Delta f C_0 w,
\end{eqnarray}
where $\sigma$ is the energy per unit length required to form a free
boundary in a crystalline bulk by breaking interparticle bonds, and
$\Delta f$ is the change in free energy of the non-crystalline phase
relative to the crystalline phase \cite{chaikin1995principles}. Note
that $\Delta f > 0$ is positive, as the disordered phase has higher
energy than the ordered crystalline phase. Since the energy in
Eq.~\ref{eq:Ecylinder} decreases with increasing crystal width $w$ for
any $w > 0$, crystal growth always lowers the energy of the system.
Therefore, the seeded crystal would nucleate and grow to cover the
entire cylindrical surface. Note that the lack of a nucleation barrier
here is specific to our ring seeding configuration. An energy barrier is
typically present for nucleation from a more isotropic seed on a
two-dimensional surface.

In contrast, on the surface of a cone, crystal growth from an initially
unstrained ring of particles can generate nonzero elastic strain. In the
limit of a small cone angle, a crystalline band of width $w$ centered
about an initial commensurate ring of circumference $C_0$ has
energy~\cite{barber2002elasticity}
\begin{equation}
    E_\mathrm{cone} = E_\mathrm{bound} + E_\mathrm{bulk} + E_\mathrm{strain},
\end{equation}
where
\begin{eqnarray}
E_\mathrm{bound} &=& 2  \sigma C_0. \label{eq:Ebound} \\
E_\mathrm{bulk} 
&=& - \Delta f C_0 w  \label{eq:Ebulk} \\
E_\mathrm{strain} &=&\frac{ Y \tan^2 (\pi \sin \beta/2)}{6 C_0} w^3,\label{eq:Estrain}
\end{eqnarray}
where $\beta$ is the cone angle defined in Figs.~\ref{fig:geometry}
and~\ref{fig:ring} and $Y$ is the two-dimensional Young's modulus of the
colloidal crystal. Owing to the $\sim w^3$ scaling of the strain energy
in Eq.~\ref{eq:Estrain}, the derivative of the total energy $E$ with
respect to $w$ becomes positive above a critical width $w_c$ given by
\begin{eqnarray} \label{eq:wc}
w_c = \frac{2 C_0}{\pi \beta} \sqrt{\frac{2 \Delta f}{Y} },
\end{eqnarray}
indicating the width at which colloidal crystallization is arrested.
Note that the nonzero Gaussian curvature of the conic surface, which
manifests as the varying circumference of the cone cross-section, is an
essential ingredient for this growth arrest.

Interestingly, we can also leverage this crystallization arrest
phenomenon to propose an alternative method for measuring the Young's
modulus of the colloidal crystal, a quantity that has typically been
extracted through various inference methods~\cite{mellor2005probing,
  meng_elastic_2014}. On inverting Eq.~\ref{eq:wc}, we have
\begin{eqnarray} \label{eq:Y}
Y = 8 \Delta f\left(\frac{ C_0}{\pi \beta w_c} \right)^2.
\end{eqnarray}
Since the critical width of a crystalline band on a cone is a directly
measurable quantity, we can use it to infer the Young's modulus $Y$
through Eq.~\ref{eq:Y}.

In this calculation, and throughout the paper, we have assumed that the
triangular lattice is the most stable configuration. For longer-ranged
interactions, however, it is possible that the crystal could restructure
itself to accommodate the elastic strain. Future work might examine
whether other lattices, such as the rhombic lattice studied by Mughal
and Weaire~\citep{mughal_theory_2014} in the context of disk packings on
cylinders, might represent stable structures on the cone.

\end{document}